\documentclass[aps,prd,superscriptaddress,showpacs,showkeys,floatfix,twocolumn,nofootinbib]{revtex4}

\usepackage{amsmath,amssymb,amsfonts,dcolumn,mathcomp,slashed,dsfont}
\usepackage{graphicx}
\usepackage{citesort}
\usepackage{psfrag}

\def\Xint#1{\mathchoice
   {\XXint\displaystyle\textstyle{#1}}%
   {\XXint\textstyle\scriptstyle{#1}}%
   {\XXint\scriptstyle\scriptscriptstyle{#1}}%
   {\XXint\scriptscriptstyle\scriptscriptstyle{#1}}%
   \!\int}
\def\XXint#1#2#3{{\setbox0=\hbox{$#1{#2#3}{\int}$}
     \vcenter{\hbox{$#2#3$}}\kern-0.5\wd0}}
\newcommand{\dashint}[1]{\Xint{\hspace{#1}-}}

\renewcommand{\Im}{\text{Im}\,}

\newcommand{\mpi}{M_{\pi}}
\newcommand{\mpii}{M_{\pi^0}}
\newcommand{\Fpi}{F_\pi}
\newcommand{\beq}{\begin{equation}}
\newcommand{\eeq}{\end{equation}}
\newcommand{\eps}{\epsilon}
\newcommand{\Order}{\mathcal{O}}
\newcommand{\F}{\mathcal{F}}
\newcommand{\G}{\mathcal{G}}
\newcommand{\diff}{\text{d}}
\newcommand{\sm}{s_{\rm m}}
\providecommand{\MeV}{\,\text{MeV}}
\providecommand{\GeV}{\,\text{GeV}}

\begin{document}

\title{Extracting the chiral anomaly from \boldmath{$\gamma\pi\to\pi\pi$}}

\author{Martin Hoferichter}
\email{hoferichter@itp.unibe.ch}
\affiliation{Helmholtz-Institut f\"ur Strahlen- und Kernphysik (Theorie) and\\
             Bethe Center for Theoretical Physics,
             Universit\"at Bonn,
             D-53115 Bonn, Germany}
\affiliation{Albert Einstein Center for Fundamental Physics, Institute for Theoretical Physics,
	    Universit\"at Bern, Sidlerstrasse 5, 
	    CH-3012 Bern, Switzerland}

\author{Bastian Kubis}
\email{kubis@hiskp.uni-bonn.de}
\affiliation{Helmholtz-Institut f\"ur Strahlen- und Kernphysik (Theorie) and\\
             Bethe Center for Theoretical Physics,
             Universit\"at Bonn,
             D-53115 Bonn, Germany}

\author{Dimitrios Sakkas}
\email{sakkas@hiskp.uni-bonn.de}
\affiliation{Helmholtz-Institut f\"ur Strahlen- und Kernphysik (Theorie) and\\
             Bethe Center for Theoretical Physics,
             Universit\"at Bonn,
             D-53115 Bonn, Germany}

%\date{\today}

\begin{abstract}
We derive dispersive representations for the anomalous process $\gamma\pi\to\pi\pi$ with the $\pi\pi$ $P$-wave phase shift as input. We investigate how in this framework the chiral anomaly can be extracted from a cross-section measurement using all data up to $1\GeV$, and discuss the importance of a precise representation of the  $\gamma\pi\to\pi\pi$ amplitude for the hadronic light-by-light contribution to the anomalous magnetic moment of the muon.  
\end{abstract}

\pacs{11.55.Fv, 13.75.Lb, 11.30.Rd, 13.60.Le}

\keywords{Dispersion relations, Meson--meson interactions, Chiral Symmetries, Meson production}

\maketitle

\section{Introduction}

The low-energy dynamics of the process $\gamma\pi\to\pi\pi$ are governed by the Wess--Zumino--Witten anomaly~\cite{WZW}. At leading order in the chiral expansion the amplitude is fully determined by the electric charge $e$, the pion decay constant $\Fpi=(92.21\pm 0.14)\MeV$~\cite{PDG}, and the number of colors $N_c$~\cite{WZW,LET}
\beq
\label{LET_3pi}
F_{3\pi}=\frac{e N_c}{12\pi^2 \Fpi^3}=(9.78\pm 0.05)\GeV^{-3}.
\eeq 
While the analogous low-energy theorem for the neutral pion decay $\pi^0\to\gamma\gamma$~\cite{LET_pi0},
\beq
\label{LET_2pi}
F_{\pi\gamma\gamma}=\frac{e^2 N_c}{12\pi^2 \Fpi},
\eeq 
concords with experiment to a remarkable accuracy (see Ref.~\cite{Bernstein_review} for a recent review), agreement between the chiral prediction~\eqref{LET_3pi} and experimental data has long proven elusive. Information on $F_{3\pi}$ can be extracted from a Primakoff reaction where a charged pion scatters off the Coulomb field of a heavy nucleus, giving access to the cross section for $\gamma\pi^-\to\pi^-\pi^0$ and thus to the chiral anomaly. Assuming that the amplitude were solely given by Eq.~\eqref{LET_3pi}, the Primakoff experiment in the threshold region performed 
at Serpukhov~\cite{Antipov} led to $F_{3\pi}=(12.9\pm 0.9\pm 0.5)\GeV^{-3}$, suggesting some tension with the low-energy theorem. 

Subsequently, it was shown that neither the one-loop~\cite{Bijnens90} and two-loop~\cite{Hannah} corrections, nor the inclusion of vector mesons~\cite{Holstein} can fully resolve this discrepancy, while a treatment based on dispersion relations was found to suffer from the occurrence of unknown free parameters~\cite{Truong}. A study of leading-logarithm contributions shows negligible corrections beyond one loop~\cite{LeadingLogs}. For theoretical approaches based on quark-loop and Dyson--Schwinger-type calculations, see Ref.~\cite{Benic} and references therein. Finally, large electromagnetic corrections in $\gamma\pi^-\to\pi^-\pi^0$ predominantly due to the $t$-channel exchange of a virtual photon were discovered in Ref.~\cite{Ametller}, which in combination with higher-order chiral corrections reduce the value of the inferred chiral anomaly to $F_{3\pi}=(10.7\pm 1.2)\GeV^{-3}$ and thus remove the tension with the low-energy theorem. Later on, the prediction for the chiral anomaly was also tested in 
$\pi^- e^-\to \pi^- e^- \pi^0$~\cite{Scherer}, leading to an extracted value of $F_{3\pi}=(9.6\pm 1.1)\GeV^{-3}$. 
Both results are now compatible with the theoretical prediction, which, however, is only tested at the $10\%$ level. Therefore, it would clearly be desirable to improve this accuracy and confront the low-energy theorem using better data and a refined theoretical approach. 

Presently, a Primakoff measurement of $\gamma\pi^-\to\pi^-\pi^0$ is being analyzed at COMPASS, with the aim of determining the cross section not only in the low-energy regime as in the Serpukhov experiment~\cite{Antipov}, but also for center-of-mass energies well beyond, in particular including the peak generated by the $\rho(770)$ resonance~\cite{Nagel}. However, the chiral amplitude will only be valid in the low-energy region, so that most of the data would be lost for the anomaly extraction.
In this paper we propose a dispersive framework that can be used to deduce the chiral anomaly from a fit to the full data set up to roughly $1\GeV$, incorporating the physics of the $\rho(770)$ by means of the $\pi\pi$ $P$-wave phase shift, and thus vastly improving the statistical accuracy of the resulting anomaly determination. Moreover, we carefully revisit the assumptions required to perform the extrapolation to the chiral limit and study the internal consistency of our dispersive framework by comparing representations with a finite and an infinite matching point. 
  
Apart from the test of chiral dynamics renewed interest in $\gamma\pi\to\pi\pi$ has been triggered recently by its impact on hadronic light-by-light scattering, whose uncertainty may soon dominate the error budget in the theoretical prediction of the anomalous magnetic moment of the muon (see Ref.~\cite{g-2:review} for a review). More precisely, one of the most important contributions, the $\pi^0$ pole term, will be determined by the doubly-virtual decay $\pi^0\to\gamma^*\gamma^*$ and thus the corresponding form factor $F_{\pi^0\gamma^*\gamma^*}(\mpii^2,q_1^2,q_2^2)$, where $q_{1/2}^2$ refer to the photon virtualities. For one of the photons being on-shell, i.e., the process $\pi^0\to\gamma\gamma^*$, one may again write down a dispersion relation based on the two-pion cut, and the $\gamma\pi\to\pi\pi$ amplitude serves as vital input for such a representation. (See Ref.~\cite{Schneider2012} for the same arguments in the context of the $\omega,\phi\to\pi^0\gamma^*$ transition form factors.)

This article is organized as follows: in Sec.~\ref{sec:kinematics}, we first introduce the notation and conventions necessary for the dispersive representations for $\gamma\pi\to\pi\pi$ that we derive and compare in Sec.~\ref{sec:disp_rep}. The consequences for the extraction of the chiral anomaly and the relation to the muon anomalous magnetic moment are discussed in Secs.~\ref{sec:anomaly} and \ref{sec:g-2}, before we close with a summary in Sec.~\ref{sec:summary}.

\section{Kinematics and partial-wave decomposition}
\label{sec:kinematics}

We decompose the amplitude for the process
\beq
\gamma(q)\pi^-(p_1)\to\pi^-(p_2)\pi^0(p_0)
\eeq
in terms of the scalar function $\F(s,t,u)$ according to
\beq
\mathcal{M}(s,t,u)=i\eps_{\mu\nu\alpha\beta}\eps^\mu p_1^\nu p_2^\alpha p_0^\beta \F(s,t,u)
\eeq
and Mandelstam variables chosen as $s=(q+p_1)^2$, $t=(p_1-p_2)^2$, and $u=(p_1-p_0)^2$. Working in the isospin limit with $\mpi=M_{\pi^+}$, we have on the mass shell  $s+t+u=3\mpi^2$, and in the center-of-mass frame we may write 
\begin{align}
\label{kin_t}
t&=a_s+b_s z,\quad u=a_s-b_s z,\\
a_s&=\frac{3\mpi^2-s}{2},\quad b_s=\frac{s-\mpi^2}{2}\sigma^\pi_s,\quad \sigma^\pi_s=\sqrt{1-\frac{4\mpi^2}{s}},\notag
\end{align}
with scattering angle $z=\cos\theta$. By virtue of isospin symmetry, the scalar function $\F(s,t,u)$ is fully symmetric in its arguments. The partial-wave decomposition then takes the form~\cite{JW}
\beq
\F(s,t,u)=\sum\limits_{\text{odd }l}f_l(s)P_l'(z),
\eeq
where $P_l'(z)$ denotes the derivative of the Legendre polynomials. In our analysis we will only be concerned with the $P$-wave $f_1(s)$, whose projection formula reads
\beq
f_1(s)=\frac{3}{4}\int^1_{-1}\diff z\big(1-z^2\big)\F(s,t,u).
\eeq
In the absence of inelastic contributions, its imaginary part is given by
\beq
\label{unitarity}
\Im f_1(s)=\sigma^\pi_s\big(t^1_1(s)\big)^*f_1(s)\theta\big(s-4\mpi^2\big),
\eeq
with the $\pi\pi$ $P$-wave amplitude
\beq
t^1_1(s)=\frac{e^{2i\delta^1_1(s)}-1}{2i\sigma^\pi_s}
\eeq
parameterized in terms of the phase shift $\delta^1_1(s)$. The unitarity relation~\eqref{unitarity} immediately implies Watson's final-state theorem~\cite{Watson}, namely that the phase of $f_1(s)$ coincides with $\delta^1_1(s)$. Finally, the formula for the cross section 
\beq
\sigma(s)=\frac{\big(s-4\mpi^2\big)^{3/2}\big(s-\mpi^2\big)}{1024\pi\sqrt{s}}\int^1_{-1}\diff z\big(1-z^2\big)|\F(s,t,u)|^2
\eeq
in the isospin limit should be augmented by 
\beq
\label{em_corr}
\F(s,t,u)\to\F(s,t,u)-\frac{2e^2\Fpi^2}{t}F_{3\pi}
\eeq
in order to include the dominant electromagnetic correction~\cite{Ametller}. Strictly speaking, this correction depends on the anomalous $\pi^0\to\gamma\gamma^*$ form factor as well as the pion vector form factor $F_\pi^V(t)$. However, higher-order terms in the radiative corrections will be of minor importance, so that the leading form of the chiral prediction, see Eqs.~\eqref{LET_3pi} and \eqref{LET_2pi}, which has been employed in Eq.~\eqref{em_corr}, should prove adequate. Likewise, subleading electromagnetic corrections were shown to be numerically irrelevant in Ref.~\cite{Ametller}.\footnote{One might be tempted to think that $t$-channel $\omega$
exchange should also yield a significant isospin-breaking
contribution, given the enhanced $\omega\pi^0\gamma$ coupling.  In a
hypothetical measurement of $\gamma\pi^0\to\pi^+\pi^-$, this would be a
very sizable effect, enhanced compared to the $\rho$-$\omega$ mixing
signal in the pion vector form factor by about a factor of $9$.
However, as a $t$-channel contribution, it is suppressed at the
sub-percent level even in the threshold region for
$\gamma\pi^-\to\pi^-\pi^0$, and becomes entirely insignificant as soon as the energy increases.}

\section{Dispersive representations}
\label{sec:disp_rep}

In any application of dispersion relations one of the basic assumptions concerns the behavior of the spectral function at high energies, i.e., the convergence properties of the dispersive integral. As elastic unitarity only determines the imaginary part below the onset of inelastic channels, it is crucial that the uncertainties associated with the high-energy input be carefully investigated. In the infinite-matching-point setup these uncertainties are reflected in the behavior of the phase shifts at high energies as well as the contributions from inelastic channels. It then needs to be shown that the final results are insensitive to these assumptions, which will be fulfilled provided the integral converges sufficiently fast. In contrast, in the finite-matching-point formulation the phase shifts are needed only in a finite energy domain, while the imaginary part above the so-called matching point $\sm$ is taken as input. The uncertainties generated by the input for the spectral function above $\sm$ correspond 
to the uncertainties induced by the high-energy region in the dispersive integral in the infinite-matching-point case. In this section we will derive and compare both approaches for $\gamma\pi\to\pi\pi$.

\subsection{Infinite matching point and angular averages}

Neglecting the imaginary parts of partial waves with angular momentum $l\geq 3$ the amplitude $\F(s,t,u)$ may be decomposed as (cf.\ Ref.~\cite{Niecknig})
\beq
\label{decomp}
\F(s,t,u)=\F(s)+\F(t)+\F(u).
\eeq
In the spirit of the ``reconstruction theorem'' in the context of chiral perturbation theory (ChPT)~\cite{reconstruction}, it can be shown that this decomposition holds exactly up to corrections of chiral power $\Order(p^{10})$.
The generalization of this representation including the absorptive part of the $F$-wave can be found in Ref.~\cite{Niecknig}, but in this article Eq.~\eqref{decomp} will be sufficient. $\F(s,t,u)$ fulfills the once- and twice-subtracted dispersion relations~\cite{Hannah}
\begin{align}
\label{def}
 &\F(s,t,u)\\
&\!=C_1+\frac{1}{\pi}\int_{4\mpi^2}^\infty\frac{\diff s'}{s'}\bigg\{\frac{s}{s'-s}+\frac{t}{s'-t}+\frac{u}{s'-u}\bigg\}\Im f_1(s'),\notag\\
&\!=C_2+\frac{1}{\pi}\int_{4\mpi^2}^\infty\frac{\diff s'}{s'^2}\bigg\{\frac{s^2}{s'-s}+\frac{t^2}{s'-t}+\frac{u^2}{s'-u}\bigg\}\Im f_1(s').\notag
\end{align}
By virtue of Eq.~\eqref{decomp}, these dispersion relations correspond to
\begin{align}
\label{disp_rel}
\F(s)&=\frac{C_1}{3}+\frac{1}{\pi}\int_{4\mpi^2}^\infty\frac{\diff s'}{s'}\frac{s}{s'-s}\Im \F(s'),\\
&=\frac{1}{3}\big(C_2^{(1)}+C_2^{(2)} s\big)+\frac{1}{\pi}\int_{4\mpi^2}^\infty\frac{\diff s'}{s'^2}\frac{s^2}{s'-s}\Im \F(s'),\notag
\end{align}
with
\beq
\label{twice_sub_const}
C_2^{(1)}+C_2^{(2)} \mpi^2=C_2.
\eeq
In addition, we define
\beq
\label{Pwave}
f_1(s)=\F(s)+\hat \F(s),
\eeq
with the hat function
\begin{align}
\label{hat_def}
\hat \F(s)&=3\big\langle \big(1-z^2\big)\F\big\rangle,\quad \langle z^n\F\rangle=\frac{1}{2}\int_{-1}^1\diff z z^n \F(t),
\end{align}
being real on the right-hand cut,
so that due to elastic unitarity~\eqref{unitarity},
\begin{align}
\Im f_1(s)&=\Im \F(s)\\
&=\big(\F(s)+\hat \F(s)\big)\theta\big(s-4\mpi^2\big)\sin \delta^1_1(s) e^{-i\delta^1_1(s)}.\notag
\end{align}
The solution of this equation for the once- and twice-subtracted versions becomes~\cite{MuskOmnes,AnLeut}
\begin{align}
 \F(s)&=\Omega(s)\Bigg\{\frac{C_1}{3}+\frac{s}{\pi}\int_{4\mpi^2}^\infty\diff s'\frac{\hat \F(s')\sin\delta^1_1(s')}{s'(s'-s)|\Omega(s')|}\Bigg\}\notag\\
&=\Omega(s)\Bigg\{\frac{C_2^{(1)}}{3}\big(1-\dot\Omega(0)s\big)+\frac{C_2^{(2)}}{3}s \label{FOmnes}\\
&\qquad\qquad\qquad+\frac{s^2}{\pi}\int_{4\mpi^2}^\infty\diff s'\frac{\hat \F(s')\sin\delta^1_1(s')}{s'^2(s'-s)|\Omega(s')|}\Bigg\},\notag
\end{align}
with the Omn\`es function
\beq
\Omega(s)=\exp\Bigg\{\frac{s}{\pi}\int_{4\mpi^2}^\infty\diff s'\frac{\delta^1_1(s')}{s'(s'-s)}\Bigg\}
\eeq
and its derivative $\dot\Omega(s)$.
In combination with Eq.~\eqref{hat_def}, these relations allow for an iterative calculation of $\hat \F(s)$, which, in turn, determines $f_1(s)$ by means of Eq.~\eqref{Pwave}. One may check explicitly that this representation fulfills Watson's theorem, and eventually leads to
\begin{align}
\label{sol_infinite}
|f_1(s)|&=\hat \F(s)\cos\delta^1_1(s)\notag\\
&+|\Omega(s)|\Bigg\{\frac{C_1}{3}+\frac{s}{\pi}\dashint{0.5pt}_{4\mpi^2}^\infty\diff s'\frac{\hat \F(s')\sin\delta^1_1(s')}{s'(s'-s)|\Omega(s')|}\Bigg\}\notag\\
&=\hat \F(s)\cos\delta^1_1(s)+|\Omega(s)|\Bigg\{\frac{C_2^{(1)}}{3}\big(1-\dot\Omega(0)s\big)\notag\\
&+\frac{C_2^{(2)}}{3}s+\frac{s^2}{\pi}\dashint{0.5pt}_{4\mpi^2}^\infty\diff s'\frac{\hat \F(s')\sin\delta^1_1(s')}{s'^2(s'-s)|\Omega(s')|}\Bigg\},
\end{align}
where the dash denotes the principal value of the integral.
The full amplitude can then be reconstructed via Eqs.~\eqref{def}, \eqref{twice_sub_const}, and Watson's theorem.

\subsection{Finite matching point and kernel functions}

The key point of the framework presented in the previous subsection concerns the fact that the calculation of the angular averages in Eq.~\eqref{hat_def}
 has to be done numerically. In an alternative approach, which proves convenient for the finite-matching-point formulation, first the original dispersive representation~\eqref{def} is inserted again into Eq.~\eqref{hat_def}, the angular integral is performed analytically, and only in the last step is the final solution expressed in terms of Omn\`es functions (for details of this procedure we refer to Ref.~\cite{RS}), so that in the end the angular integration becomes traded off in favor of another dispersive integral. In this way, the result for the modulus,
\begin{align}
\label{sol_finite}
 |f_1(s)|&=\Delta_1(s)\cos\delta^1_1(s)\notag\\
&\qquad+|\Omega(s)|\Bigg\{C_1+\frac{s}{\pi}\dashint{0.5pt}_{4\mpi^2}^{\sm}\diff s'\frac{\Delta_1(s')\sin\delta^1_1(s')}{s'(s'-s)|\Omega(s')|}\notag\\
&\qquad+\frac{s}{\pi}\int\limits_{\sm}^\infty\diff s'\frac{\Im f_1(s')}{s'(s'-s)|\Omega(s')|}\Bigg\}\notag\\
&=\Delta_2(s)\cos\delta^1_1(s)+|\Omega(s)|\Bigg\{C_2\big(1-\dot\Omega(0)s\big)\notag\\
&\qquad+\frac{s^2}{\pi}\dashint{0.5pt}_{4\mpi^2}^{\sm}\diff s'\frac{\Delta_2(s')\sin\delta^1_1(s')}{s'^2(s'-s)|\Omega(s')|}\notag\\
&\qquad+\frac{s^2}{\pi}\int_{\sm}^\infty\diff s'\frac{\Im f_1(s')}{s'^2(s'-s)|\Omega(s')|}\Bigg\}
\end{align}
indeed resembles Eq.~\eqref{sol_infinite}, but now the inhomogeneities $\Delta_n(s)$, $n\in\{1,2\}$, are not given by an angular integral, but by 
\beq
\Delta_n(s)=\frac{1}{\pi}\int_{4\mpi^2}^\infty\diff s'K_n(s,s')\Im f_1(s'),
\eeq
with kernel functions
\begin{align}
 K_1(s,s')&=\frac{3}{b_s}\Big\{\big(1-x_s^2\big)Q_0(x_s)+x_s\Big\}-\frac{2}{s'},\notag\\
K_2(s,s')&=K_1(s,s')+\frac{s-3\mpi^2}{s'^2},\quad x_s=\frac{s'-a_s}{b_s},
\end{align}
where
\begin{align}
Q_0(z)&=\frac{1}{2}\int^1_{-1}\frac{\diff x}{z-x},\notag\\
Q_0(z\pm i\eps)&=\frac{1}{2}\log\bigg|\frac{1+z}{1-z}\bigg|\mp i\frac{\pi}{2}\theta\big(1-z^2\big),
\end{align}
denotes the lowest Legendre function of the second kind. In addition, we have introduced a finite matching point $\sm$; i.e., $\Im f_1(s)$ is assumed to be known for $s\geq \sm$. The Omn\`es function
\beq
\Omega(s)=\exp\Bigg\{\frac{s}{\pi}\int_{4\mpi^2}^{\sm}\diff s'\frac{\delta^1_1(s')}{s'(s'-s)}\Bigg\}
\eeq
now only involves the phase shift below $\sm$, but exhibits a cusp in the vicinity of $\sm$,
\beq
\label{cusp}
\Omega(s)\sim |s-\sm|^x,\quad x=\frac{\delta^1_1(\sm)}{\pi},
\eeq
which complicates the numerical implementation, cf.\ Ref.~\cite{RS}.
 Once this is accounted for, Eq.~\eqref{sol_finite} provides an iterative scheme that determines $|f_1(s)|$ for $4\mpi^2\leq s\leq \sm$, and thus, by means of Eq.~\eqref{def}, the full amplitude $\F(s,t,u)$.\footnote{The dispersive treatment in Ref.~\cite{Truong} employs this kernel-based approach in the infinite-matching-point case. Since the behavior of the inhomogeneities $\Delta_n(s)$ for large $s$ induces a rather slow convergence of the (second) dispersive integral for $s\to\infty$, we only use this formulation in combination with a finite matching point.} 

Another interesting feature of the finite-matching-point derivation concerns the fact that statements about the strict validity of the final integral equation in the context of the convergence of the partial-wave expansion are possible. Given certain analyticity properties of $\F(s,t,u)$, one may calculate the corresponding maximally allowed value of the matching point; e.g., starting from fixed-$t$ dispersion relations and assuming Mandelstam analyticity~\cite{Mandelstam}, we obtain
\beq
s_\text{max}=(1.2\GeV)^2,
\eeq  
which already indicates that a dispersive representation cannot be rigorously valid far above $1\GeV$. Further details of the derivation of $s_\text{max}$ are collected in Appendix~\ref{app:range_of_conv}.

\subsection{Comparison}

The free parameters of the dispersive representations presented in the previous subsections will be determined by fitting to the experimental result for the cross section $\sigma(s)$ once available. With two subtractions, these are the constants $C^{(1)}_2$ and $C^{(2)}_2$ in the infinite-matching-point case and $C_2$ for a finite matching point. Moreover, absent independent information on the imaginary part, $\Im f_1(s)$ will simply be set equal to zero above $\sm$,\footnote{Indeed, the (preliminary, uncorrected) count rates presented in Ref.~\cite{Nagel} indicate that at high energies the cross section becomes essentially zero.  If the final cross section at high energies turned out to be non-negligible, the corresponding effect could be included by modeling the high-energy tail of $f_1$ in such a way that the experimental cross section be reproduced. Either way, for a sufficiently high matching point the influence on the energy region $\lesssim 1\GeV$ is expected to be of minor importance, cf.\ Ref.~\cite{RS}.} so that $\sm$ can be regarded as another free parameter that determines at what energy the amplitude is guided to zero. From this point of view, both representations involve two free parameters that need to be fitted to experiment.

As a starting point for the numerical implementation we take $\F(s,t,u)=0$, but the outcome of the iteration is completely insensitive to this choice. (We have checked that e.g.\ using a vector-meson-dominance ansatz as starting point leads to exactly the same results.) The solutions for $|f_1(s)|$ and $\hat \F(s)$, for the finite- and infinite-matching-point formulations, respectively, are obtained by iterating the integral equation until these functions change by less than $10^{-5}$, which, for the infinite matching point, requires six iterations. We find that the iterative scheme for the infinite matching point converges faster than its finite-matching-point equivalent: the number of iterations needed for a given accuracy is reduced by roughly a factor $2$, and the complications due to the cusp of the finite-matching-point Omn\`es function~\eqref{cusp} are absent. On the other hand, for the implementation of the infinite-matching-point scheme, one needs to specify an integration cutoff $\Lambda$ and study the sensitivity to a particular choice of $\Lambda$. In the following, 
we take $\Lambda=1.8\GeV$, which proves sufficiently large for the energy region $\lesssim 1\GeV$ to remain unaffected under cutoff variations.    

\begin{figure}
\centering
\includegraphics[width=\linewidth, clip]{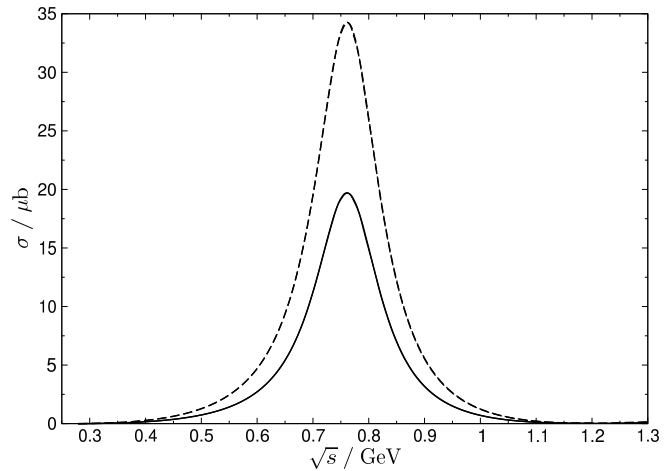}
\caption{Cross section for $\gamma\pi\to\pi\pi$ for $C_2=9.78\GeV^{-3}$ (solid line) and $C_2=12.9\GeV^{-3}$ (dashed line). The curves for the finite/infinite-matching-point setup lie on top of each other.}
\label{fig:cross_section}
\end{figure}

For illustrative purposes, we now take $\sm=(1.2\GeV)^2$ and fix $C^{(2)}_2$ in such a way that for a given $C_2$ the infinite-matching-point amplitude vanishes at $\sm$. 
We include the electromagnetic corrections according to Eq.~\eqref{em_corr}, although these are only relevant in the low-energy region where the cross section is rather small (see Fig.~2 in Ref.~\cite{Ametller}), so that these corrections mainly affect the conversion between $\sigma(s)$ and the total cross section of the Primakoff reaction~\cite{Ametller}. Taking $\delta^1_1(s)$ from the analysis of Roy and Roy-like equations in Refs.~\cite{CCL,GKPY}, we obtain the cross section depicted in Fig.~\ref{fig:cross_section}, where, as expected given the choice of $C^{(2)}_2$, the finite/infinite-matching-point results are indistinguishable on the scale chosen. The only difference concerns the behavior above $\sm$, where the finite-matching-point curve stays at zero, whereas the infinite-matching-point cross section slowly starts to rise again. This example nicely demonstrates the inherent consistency between the two methods, and we expect that as soon as data are available, the comparison between both fits should 
allow for a valuable consistency check regarding the extracted value of the chiral anomaly. Moreover, Fig.~\ref{fig:cross_section} shows that already moderate changes in $C_2$ entail large differences in the height of the $\rho$ peak, so we expect an enhanced sensitivity to the chiral anomaly if indeed the whole spectrum is included in the analysis.

\section{Extracting the chiral anomaly from cross-section data}
\label{sec:anomaly}

The extraction of the chiral anomaly from experimental cross-section data will proceed in two steps: first, the free parameters of the dispersive representation(s) of the previous section, the subtraction constants, need to be fitted to data; second, these subtraction constants have to be related to the chiral anomaly by matching to ChPT.  We describe these two steps in the following subsections.

\subsection{Fitting dispersive representations to data}
\label{sec:DRdata}

For the practical purpose of a fit to data, the representations~\eqref{FOmnes} can be significantly simplified, based on the observation that they are completely \emph{linear} in the subtraction constants $C_1$ and $C_2^{(1)}$, $C_2^{(2)}$, respectively~\cite{Niecknig,BernPC}.  We can rewrite
\beq
\F(s) = C_1 \F_1(s) = C_2^{(1)}\F_2^{(1)}(s) +  C_2^{(2)}\F_2^{(2)}(s) , \label{linear} 
\eeq
with
\begin{align}
\F_1(s)&=\Omega(s)\Bigg\{\frac{1}{3}+\frac{s}{\pi}\int_{4\mpi^2}^\infty\diff s'\frac{\hat \F_1(s')\sin\delta^1_1(s')}{s'(s'-s)|\Omega(s')|}\Bigg\} ,\notag\\
\F_2^{(1)}(s)&=\Omega(s)\Bigg\{\frac{1-\dot\Omega(0)s}{3} \label{basis}\\
&\qquad\qquad+\frac{s^2}{\pi}\int_{4\mpi^2}^\infty\diff s'\frac{\hat \F_2^{(1)}(s')\sin\delta^1_1(s')}{s'^2(s'-s)|\Omega(s')|}\Bigg\},\notag \\
\F_2^{(2)}(s)&=\Omega(s)\Bigg\{\frac{s}{3}+\frac{s^2}{\pi}\int_{4\mpi^2}^\infty\diff s'\frac{\hat \F_2^{(2)}(s')\sin\delta^1_1(s')}{s'^2(s'-s)|\Omega(s')|}\Bigg\}, \notag
\end{align}
where $\hat\F_1(s)$, $\hat\F_2^{(1)}(s)$, and $\hat\F_2^{(2)}(s)$ have been rescaled accordingly and are all determined from $\F_1(s)$, $\F_2^{(1)}(s)$, and $\F_2^{(2)}(s)$ according to the angular averaging prescription~\eqref{hat_def}. Obviously, the \emph{basis functions} as given in Eq.~\eqref{basis} can be calculated iteratively, independently of any subtraction constants, so they can easily be provided to experimental analyses in numerical form, allowing for a straightforward extraction of the subtraction constants given as linear prefactors in Eq.~\eqref{linear}.  Furthermore, Eq.~\eqref{linear} explains the strong dependence of the cross sections shown in Fig.~\ref{fig:cross_section} on the subtraction constants: they really serve as multiplicative constants in the whole (elastic) energy range, yielding the desired strong impact of the $\rho$ resonance region on the extraction of the anomaly.

We finally remark that for the finite-matching-point scenario Eq.~\eqref{sol_finite}, a similar linearity property can be used as long as the input above the matching point is set to zero (as we have done above). For varying input above the matching point, as well as for different matching points, different basis functions would have to be calculated.

\subsection{Matching to chiral perturbation theory}
\label{sec:match}

The chiral anomaly is defined as the value of $\F(s,t,u)$ at $s=t=u=0$. Since the dispersion relations~\eqref{def} have been derived using on-shell kinematics, they cannot be used to perform the required extrapolation to the chiral limit. (For a more detailed discussion of this point see Appendix~\ref{app:offshell}.) Therefore, the appropriate procedure to extract the chiral anomaly involves on-shell matching to ChPT, with the chiral expansion thereafter used to extrapolate to the chiral limit.  

The one-loop, $\Order(p^6)$, ChPT amplitude may be expressed in the form~\cite{Bijnens90}
\begin{align}
 \F^{(6)}(s,t,u)&=\F^{(6)}(s)+\F^{(6)}(t)+\F^{(6)}(u),\notag\\
\F^{(6)}(s)&=\frac{s^2}{\pi}\int_{4\mpi^2}^\infty\diff s'\frac{\Im f_1^{(6)}(s')}{s'^2(s'-s)}\\
&\hspace{-35pt}+F_{3\pi}\Bigg\{\frac{1}{3}-\frac{64\pi^2}{3e}C^{\rm r}_2(\mu)s-\frac{s}{96\pi^2\Fpi^2}\bigg(1+\log\frac{\mpi^2}{\mu^2}\bigg)\Bigg\},\notag
\end{align}
where $C^{\rm r}_2(\mu)$ is an $\Order(p^6)$ low-energy constant (LEC), and $\Im f_1^{(6)}(s)$ denotes the one-loop imaginary part
\beq
\Im f_1^{(6)}(s)=\sigma^\pi_sf_1^{(4)}(s)t^{(2)}(s)=\sigma_s^\pi F_{3\pi}\frac{s-4\mpi^2}{96\pi\Fpi^2}.
\eeq
The dispersive representations~\eqref{def} may be summarized in the form
\begin{align}
 &\F^{\rm DR}(s,t,u)\\
&=C+\frac{1}{\pi}\int_{4\mpi^2}^\infty\frac{\diff s'}{s'^2}\bigg\{\frac{s^2}{s'-s}+\frac{t^2}{s'-t}+\frac{u^2}{s'-u}\bigg\}\Im f_1(s'),\notag
\end{align}
with $C=C_2$ for two subtractions and
\beq
C=C_1+\frac{3\mpi^2}{\pi}\int_{4\mpi^2}^\infty\diff s'\frac{\Im f_1(s')}{s'^2}
\eeq
for the once-subtracted case. After the fit of the dispersive amplitude to cross-section data, either in the finite- or infinite-matching-point scenario, the partial wave $f_1(s)$ as well as the constant $C$ will be known. Demanding that
\beq
\F^{\rm DR}(s,t,u)=\F^{(6)}(s,t,u)+\Order\big(p^8\big)
\eeq 
in the low-energy region shows that
\begin{align}
\label{renormalization}
C&=F_{3\pi}\big(1+3\mpi^2\bar C\big),\notag\\
\bar C&=-\frac{64\pi^2}{3e}C_2^{\rm r}(\mu)-\frac{1}{96\pi^2\Fpi^2}\bigg(1+\log\frac{\mpi^2}{\mu^2}\bigg),
\end{align}
since the difference in the integrals is of higher chiral order. In consequence, one cannot disentangle the chiral anomaly from its quark-mass renormalization without further input for $C_2^{\rm r}(\mu)$, e.g., its resonance-saturation estimate~\cite{Bijnens90}
\beq
\label{resonance}
C_2^{\rm r}(\mu)=-\frac{3e}{128\pi^2M_\rho^2},
\eeq  
where $M_\rho$ denotes the mass of the $\rho(770)$, and the renormalization scale is identified as $\mu=M_\rho$.\footnote{Note that the estimate Eq.~\eqref{resonance} implicitly makes use of the KSFR relation~\cite{KSFR}.  Allowing for deviations in the relation between the $\rho\pi\pi$ and the $\rho\gamma$ couplings (see, e.g., Ref.~\cite{Leupold}) reduces $C_2^{\rm r}(\mu)$ by up to 20\%, still within an estimated uncertainty for the anomaly extraction at the percent level. We are grateful to Stefan Leupold for pointing this out to us. Alternative approaches to estimate this low-energy constant, based on a chiral quark model~\cite{BijnensStrandberg} or a Dyson--Schwinger equation~\cite{Jiang}, seem to us rather inconclusive at present.}

Similarly, one finds at two loops~\cite{Hannah} 
\begin{align}
 \F^{(8)}(s,t,u)&=\F^{(8)}(s)+\F^{(8)}(t)+\F^{(8)}(u),\notag\\
\F^{(8)}(s)&=F_{3\pi}\bigg\{\frac{1}{3}+\bar C' s+ \bar D\, s^2\bigg\}\notag\\
&\qquad+\frac{s^3}{\pi}\int_{4\mpi^2}^\infty\diff s'\frac{\Im f_1^{(8)}(s')}{s'^3(s'-s)},\notag\\
\bar C'&=\bar C+\frac{\mpi^2}{(4\pi \Fpi)^4}\bar c_2,\notag\\
\bar D&= \frac{1}{(4\pi\Fpi)^2}\bigg\{\frac{1}{60\mpi^2}+\frac{1}{(4\pi\Fpi)^2}\bar d_2\bigg\},
\end{align}
where $\bar c_2$ and $\bar d_2$ are the new LECs at $\Order(p^{8})$. To derive the matching condition, we rewrite the dispersive representation as
\begin{align}
\F^{\rm DR}(s,t,u)&=C+C_{\rm SR}\big(s^2+t^2+u^2\big)\notag\\
&\hspace{-30pt}+\frac{1}{\pi}\int_{4\mpi^2}^\infty\frac{\diff s'}{s'^3}\bigg\{\frac{s^3}{s'-s}+\frac{t^3}{s'-t}+\frac{u^3}{s'-u}\bigg\}\Im f_1(s'),\notag\\
C_{\rm SR}&=\frac{1}{\pi}\int_{4\mpi^2}^\infty\diff s'\frac{\Im f_1(s')}{s'^3},
\end{align}
with the result that
\beq
\F^{\rm DR}(s,t,u)=\F^{(8)}(s,t,u)+\Order\big(p^{10}\big)
\eeq
implies
\beq
\label{2loop_matching}
F_{3\pi}\big(1+3\mpi^2\bar C'\big)=C,\qquad F_{3\pi}\bar D=C_{\rm SR}.
\eeq
Due to the additional LECs $\bar c_2$ and $\bar d_2$ matching at two-loop level does not allow for an unambiguous extraction of the chiral anomaly either. However, estimating these $\Order(p^{8})$ LECs by order-of-magnitude arguments, these additional constraints may prove valuable in assessing the systematic uncertainty in the anomaly extraction, in particular by comparing the results of the one- and two-loop matching procedures. 

To summarize, the chiral anomaly would be derived from the dispersive fit using Eqs.~\eqref{renormalization} and \eqref{resonance}, potentially supplemented by the two-loop matching conditions~\eqref{2loop_matching}. According to Eq.~\eqref{resonance}, $C_2^{\rm r}(\mu)$ amounts to a $4.9\%$ effect, while the chiral logarithm at $\mu=M_\rho$ generates another $1.8\%$ correction. In view of the experience in the normal-parity sector that resonance saturation works rather well for vector mesons~\cite{EGPdR}, one would expect an uncertainty in $F_{3\pi}$ due to $C_2^{\rm r}(\mu)$ at the percent level. We stress that this particular source of uncertainty is by no means a limitation inherent solely to our approach, since any dispersive framework, and of course also a fit of the ChPT expression directly to data in the low-energy region, will face the difficulty of determining the LECs pertinent to the quark-mass renormalization.

\section{Relation to the muon anomalous magnetic moment}
\label{sec:g-2}

The pion transition form factor $F_{\pi^0\gamma^*\gamma^*}(\mpi^2,q_1^2,q_2^2)$, which determines the strength of the pion-pole contribution to light-by-light scattering, may be split into components with definite isospin according to~\cite{MesonNet}
\beq
\label{TFF}
F_{\pi^0\gamma^*\gamma^*}(\mpi^2,q_1^2,q_2^2)=F_{vs}(q_1^2,q_2^2)+(q_1\leftrightarrow q_2),
\eeq
where the first/second index refers to isovector ($v$) and isoscalar ($s$) quantum numbers of the photon with momentum $q_1$/$q_2$. If we assume the isovector spectral function to be saturated by two-pion intermediate states (neglecting heavier contributions from $4\pi$, $K\bar K$ etc.), the form factor $f_{\pi^0\gamma}(s)=F_{vs}(s,0)$ fulfills the once-subtracted dispersion relation (cf.\ Ref.~\cite{Niecknig})
\beq
\label{fpi0g_sub}
f_{\pi^0\gamma}(s)=f_{\pi^0\gamma}(0)+\frac{s}{12\pi^2}\int^\infty_{4\mpi^2}\diff s'\frac{q_{\pi}^3(s')\big(F_\pi^V(s')\big)^*f_1(s')}{s'^{3/2}(s'-s)},
\eeq
where $q_{\pi}(s)=\sqrt{s/4-\mpi^2}$ and $f_{\pi^0\gamma}(0)=F_{\pi\gamma\gamma}/2$. In this way, the determination of $f_1(s)$ from $\gamma\pi\to\pi\pi$ cross-section data would help in constraining the $\pi^0\gamma^*\gamma^*$ transition form factor for on-shell isoscalar photons. Assuming an unsubtracted version of Eq.~\eqref{fpi0g_sub} would entail a sum rule for $f_{\pi^0\gamma}(0)$, which we discuss in more detail in Appendix~\ref{app:sumrule}.

Moreover, $F_{vs}$ fulfills the dispersion relation
\beq
F_{vs}(s_1,s_2)=f_{\pi^0\gamma}(s_1)+\frac{s_2}{\pi}\int^\infty_{9\mpi^2}\diff x\frac{\Im F_{vs}(s_1,x)}{x(x-s_2)},
\eeq  
which would fix the $s_2$ dependence of $F_{vs}$ if the spectral function were known, and thus, by means of Eq.~\eqref{TFF}, the whole functional form of $F_{\pi^0\gamma^*\gamma^*}$. Here, the $\gamma\pi\to\pi\pi$ input provides the subtraction function necessary to write down a subtracted version of the dispersion relation, and therefore it should appreciably improve the convergence of the dispersive integral.

\section{Summary}
\label{sec:summary}

In this article, we have shown that a rigorous description of the anomalous process $\gamma\pi\to\pi\pi$ can be extended well beyond the chiral regime up to about 1\GeV, using methods from dispersion theory.  Two different variants have been employed and shown to yield equivalent results: an infinite-matching-point formulation with two subtraction constants, and the use of a finite matching point, where one subtraction seems to be sufficient.  Matching to chiral perturbation theory allows us to relate the subtraction constants to the chiral anomaly, which can therefore be extracted from a fit of the dispersive representation to \emph{all} data below 1\GeV, including in particular the large contribution of the $\rho$ resonance.  This should allow for a vast statistical improvement of the determination of the anomaly in the upcoming analysis of data taken by the COMPASS experiment, using the Primakoff effect.

We have pointed out that a reliable determination of the $\gamma\pi\to\pi\pi$ amplitude serves as a vital input in a future dispersive analysis of the $\pi^0$ transition form factor, which in turn is an important ingredient in the hadronic light-by-light-scattering contribution to the anomalous magnetic moment of the muon.  Given its far-reaching impact, we strongly encourage a thorough experimental analysis of this process along the lines presented in this article.

\begin{acknowledgments}
We would like to thank Jan Friedrich, Thiemo Nagel, Franz Niecknig, and Sebastian Schneider for numerous useful discussions, Gilberto Colangelo for providing us with the preliminary $\pi\pi$ phase shift results of Ref.~\cite{CCL}, and Johan Bijnens, Karol Kampf, and Stefan Lanz for helpful e-mail communication.
Partial financial support by
the DFG (CRC 16, ``Subnuclear Structure of Matter''),
by the project ``Study of Strongly Interacting Matter'' (HadronPhysics3, Grant Agreement No.~283286) 
under the 7th Framework Program of the EU,
by the Bonn--Cologne Graduate School of Physics and Astronomy,
and by the Swiss National Science Foundation
is gratefully acknowledged.
The Albert Einstein Center for Fundamental Physics is supported by the
  ``Innovations- und Kooperationsprojekt C-13'' of the ``Schweizerische
  Universit\"atskonferenz SUK/CRUS.'' 
\end{acknowledgments}

\appendix

\section{Domain of validity}
\label{app:range_of_conv}

\begin{figure}
\centering
\includegraphics[width=0.8\linewidth]{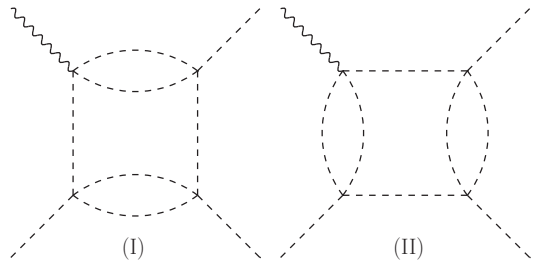}
\caption{Box graphs constraining the boundaries of the double-spectral regions.}
\label{fig:boundary}
\end{figure}

In this appendix we briefly sketch how the convergence of partial-wave expansion and projection imposes limits on the choice of the matching point. For a more thorough derivation we refer to the literature on Roy~\cite{Roy,Roy_Mandelstam} and Roy--Steiner~\cite{RS} equations. 

Mandelstam analyticity states that the scattering amplitude ${\cal T}(s,t)$ can be written in terms of double-spectral functions $\rho_{su}$, $\rho_{tu}$, and $\rho_{st}$ as
\begin{align}
{\cal T}(s,t)&=\frac{1}{\pi^2}\iint\diff s'\diff u'\frac{\rho_{su}(s',u')}{(s'-s)(u'-u)}\notag\\
&+ \frac{1}{\pi^2}\iint\diff t'\diff u'\frac{\rho_{tu}(t',u')}{(t'-t)(u'-u)}\notag\\
&+\frac{1}{\pi^2}\iint\diff s'\diff t'\frac{\rho_{st}(s',t')}{(s'-s)(t'-t)}.
\end{align}
The integration boundaries follow from the box diagrams shown in Fig.~\ref{fig:boundary}, which represent the lowest-lying intermediate states possible to contribute to the double-spectral functions. More precisely, diagrams (I) and (II) define the boundary of the support of $\rho_{st}$ as the intersection of
\begin{align}
\label{b_boundary}
b_{\text{I}}(s,t)&=8\mpi^2\big(2s+\mpi^2\big)-t\big(s-4\mpi^2\big)=0,\notag\\
b_{\text{II}}(s,t)&=4\mpi^2\big(s+2\mpi^2\big)-t\big(s-16\mpi^2\big)=0,
\end{align}
or, equivalently, by
\beq
t=T_{st}(s)=\min\big\{T_{\text{I}}(s),T_{\text{II}}(s)\big\},
\eeq  
where $T_{\text{I}}$, $T_{\text{II}}$ follow from solving Eq.~\eqref{b_boundary} for $t$.
By virtue of crossing symmetry, the equations for $\rho_{su}$ and $\rho_{tu}$ are obtained by the pertinent permutation of Mandelstam variables.

\begin{figure}
\centering
\includegraphics[width=\linewidth]{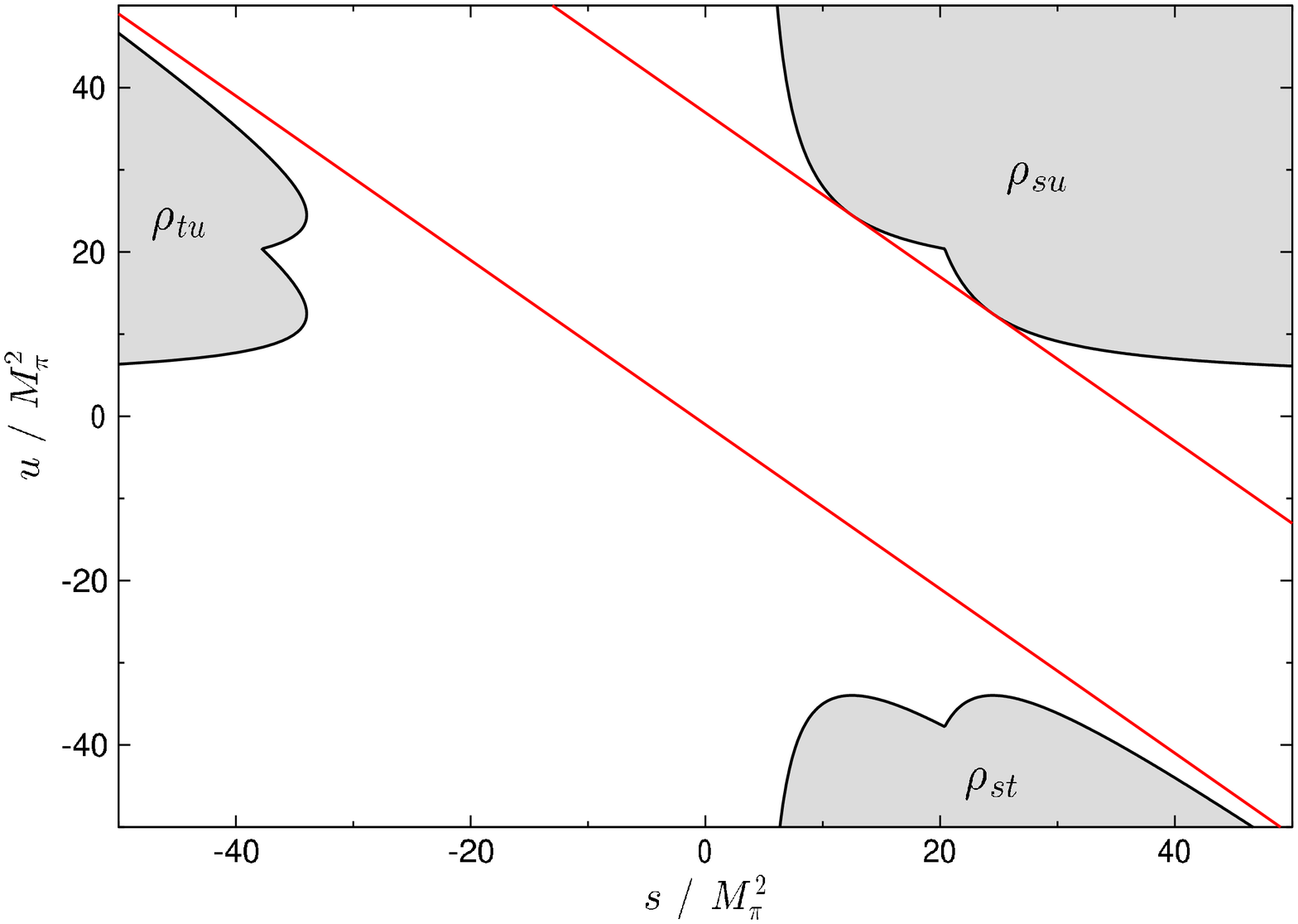}
\includegraphics[width=\linewidth]{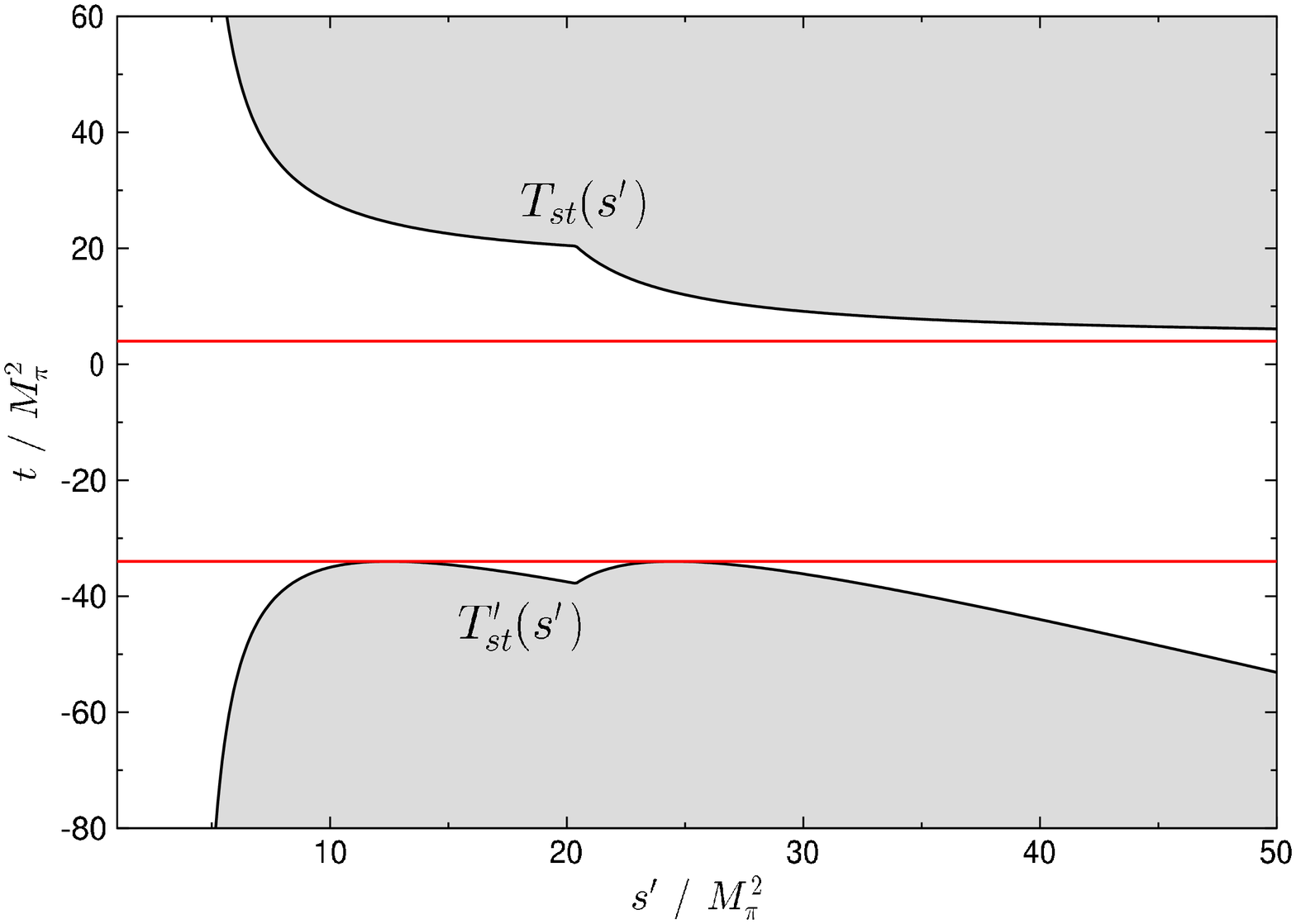}
\caption{Double-spectral regions (top) and allowed range of $t$ (bottom) for  $\gamma\pi\to\pi\pi$. The red lines refer to $t=4\mpi^2$ and $t=-34\mpi^2$, respectively.}
\label{fig:t_mandelstam_g3pi}
\end{figure}

Starting from fixed-$t$ dispersion relations,\footnote{It may be possible to further extend the range of validity by considering dispersion relations in the manifestly crossing-symmetric variables $x=st+tu+us$ and $y=stu$ instead~\cite{Roy_extended}.} these double-spectral regions limit the validity of the resulting integral equation as follows: first, the partial-wave expansion of the imaginary part inside the dispersive integral converges only for scattering angles $z$ within the large Lehmann ellipse~\cite{Lehmann58}, which is constructed as the largest ellipse in the complex $z$-plane which does not reach into the double-spectral regions. In view of Eq.~\eqref{kin_t}, this automatically constrains the range of allowed values for $t$. Second, a specific value for $t$ is only allowed if the corresponding line in the Mandelstam plane avoids the double-spectral regions as well.

The analysis of the second condition leaves the range 
\beq
\label{range_t}
-(17+12\sqrt{2})\mpi^2\approx -34\mpi^2\leq t\leq 4\mpi^2
\eeq
of allowed values for $t$ (see Fig.~\ref{fig:t_mandelstam_g3pi}), while the Lehmann-ellipse constraint can be translated into
\beq
T'_{st}(s')=3\mpi^2-s'-T_{st}(s')\leq t \leq T_{st}(s')
\eeq     
for all $s'\in[4\mpi^2,\infty)$. In fact, this condition amounts to exactly the same range already given in Eq.~\eqref{range_t} (see Fig.~\ref{fig:t_mandelstam_g3pi}).
Taking into account that by virtue of Bose symmetry we only need half the angular range to perform the partial-wave projection, i.e.
$t\in[a_s,a_s+b_s]$, we find from comparison with Eq.~\eqref{range_t}
\beq
s_\text{max}=(37+24\sqrt{2})\mpi^2\approx 71\mpi^2 \approx (1.2\GeV)^2.
\eeq

\section{Sum rule for $\boldsymbol{f_{\pi^0\gamma}(0)}$}
\label{app:sumrule}

The unsubtracted version of Eq.~\eqref{fpi0g_sub} implies the sum rule
\beq
\label{sum_rule}
f_{\pi^0\gamma}(0)=\frac{F_{\pi\gamma\gamma}}{2}=
\frac{1}{12\pi^2}\int^\infty_{4\mpi^2}\diff s'\frac{q_{\pi}^3(s')}{s'^{3/2}}\big(F_\pi^V(s')\big)^*f_1(s').
\eeq
The existence of such a sum rule can be made plausible by the following arguments. In the hidden-local-symmetry (HLS) formalism~\cite{HLS1,HLS2} with the simplest choice for the anomalous HLS couplings $c_3=c_4=1$~\cite{HLS2,HLS3} the $\gamma\pi\to\pi\pi$ amplitude reads
\begin{align}
\label{HLS}
\F(s,t,u)&=F_{3\pi}\bigg\{1+\frac{1}{2}\big(D_\rho(s)+D_\rho(t)+D_\rho(u)-3\big)\bigg\},\notag\\
D_\rho(s)&=\frac{M_\rho^2}{M_\rho^2-s}.
\end{align} 
Introducing a finite width $\Gamma_\rho$ and approximating the vector form factor in a similar fashion, the dominant (pole) part of the integrand in Eq.~\eqref{sum_rule} becomes
\beq
 \frac{e}{2} F_{3\pi}  \frac{M_\rho^4}{\big(s'-M_\rho^2\big)^2+M_\rho^2\Gamma_\rho^2}\longrightarrow \frac{e}{2} F_{3\pi} \pi \frac{M_\rho^3}{\Gamma_\rho}\delta\big(s'-M_\rho^2\big),
\eeq 
where in the last step the narrow-width approximation has been applied. Using the relation between the $\rho\pi\pi$ coupling and the partial width
\beq
\Gamma_{\rho}=\frac{g_{\rho\pi\pi}^2}{6\pi}\frac{q_\pi^3(M_\rho^2)}{M_\rho^2},
\eeq
as well as the KSFR relation $2F_\pi^2g_{\rho\pi\pi}^2=M_\rho^2$~\cite{KSFR}, we find indeed
\begin{align}
&\frac{1}{12\pi^2}\int^\infty_{4\mpi^2}\diff s'\frac{q_{\pi}^3(s')}{s'^{3/2}}\big(F_\pi^V(s')\big)^*f_1(s')\notag\\
&\hspace{50pt}\longrightarrow\frac{e}{2}F_{3\pi} F_\pi^2=\frac{F_{\pi\gamma\gamma}}{2}.
\end{align}

We can also test the sum rule with more realistic input. As an example, we use the partial wave $f_1(s)$ based on the finite-matching-point solution~\eqref{sol_finite}, with $C_2=1.066\times F_{3\pi}$ based on the chiral prediction with resonance saturation~\eqref{resonance} and $\sm=(1.2\GeV)^2$. We parameterize the pion vector form factor with a simple twice-subtracted Omn\`es representation
\beq
F_\pi^V(s)=\exp\Bigg\{\frac{\langle r^2\rangle_\pi^V}{6}s+\frac{s^2}{\pi}\int\limits_{4M_\pi^2}^{\infty}\frac{\diff s'}{s'^2}\frac{\delta_1^1(s')}{s'-s}\Bigg\},
\eeq
varying the pion charge radius in the range $\langle r^2\rangle_\pi^V=0.435\ldots0.450\,\text{fm}^2$.  Integrating the sum rule~\eqref{sum_rule} up to the matching point $\sm$ saturates the full value $F_{\pi\gamma\gamma}/2$ at $87\%\ldots90\%$, depending on the charge radius.  The remainder has to be attributed both to contributions to the integral above $\sm$ and to heavier intermediate states beyond $\pi\pi$. Given, however, the degree to which these effects (higher energies and heavier states) are suppressed already in the unsubtracted dispersion relation, we are optimistic that they should be even less important in the subtracted form suggested to study the $s$ dependence of the transition form factor in Eq.~\eqref{fpi0g_sub}.

\section{Off-shell terms and dispersion relations}
\label{app:offshell}

The dispersion relations~\eqref{def} were derived in Ref.~\cite{Hannah} starting from fixed-$t$ dispersion relations, fixing the subtraction terms by means of crossing symmetry (similarly to the derivation of $\pi\pi$ Roy equations~\cite{Roy}), and neglecting the imaginary parts of partial waves with $l\geq 3$. Although this derivation relies on on-shell kinematics $s+t+u=3\mpi^2$, the same dispersion relations were used later on to determine the quark-mass renormalization of the chiral anomaly. We will now demonstrate that this is not legitimate.

In order to obtain a version of Eq.~\eqref{def} that is valid off-shell we would have to allow for an additional term
\beq
\big(s+t+u-3\mpi^2\big)\G(s,t,u)
\eeq
with an \emph{a priori} unknown function $\G(s,t,u)$. Matching the twice-subtracted version with one-loop ChPT, we find
\beq
C_2 + \big(s+t+u-3M_\pi^2\big)\G(s,t,u) = F_{3\pi}\big(1+ \bar C(s+t+u)\big). 
\eeq
This shows that up to higher chiral orders $\G(s,t,u)\equiv \G$ is constant, and that, by comparing coefficients,
\beq
C_2 - 3M_\pi^2 \G =  F_{3\pi},\qquad
\G = F_{3\pi} \bar C, 
\eeq
which, unfortunately, merely reproduces Eq.~\eqref{renormalization}. 

The same calculation for the once-subtracted dispersion relation, now with a different function $\G'(s,t,u)$, gives
\begin{align}
 C_1 - 3M_\pi^2 \G' =  F_{3\pi},\quad \frac{1}{\pi}\int^\infty_{4\mpi^2}\diff s' \frac{\Im f_1(s')}{s'^2} + \G' = F_{3\pi} \bar C, 
\end{align}
and thus again recovers Eq.~\eqref{renormalization}. Moreover, the second equation corresponds to the sum rule for $\bar C$ from Ref.~\cite{Hannah} if $\G'$ is neglected (leading to $\bar C=0.93\GeV^{-2}$~\cite{Hannah}). Numerically, this value is rather close to~\cite{Bijnens90}
\beq
\bar C=\frac{1}{2M_\rho^2}-\frac{1}{96\pi^2\Fpi^2}\bigg(1+\log\frac{\mpi^2}{M_\rho^2}\bigg)=1.13\GeV^{-2},
\eeq 
which, however, simply indicates that the off-shell effects parameterized by $\G'$ are moderate. In particular, the HLS expression~\eqref{HLS}
\beq
\Im f_1(s)=F_{3\pi}\frac{\pi}{2}M_\rho^2\delta\big(s-M_\rho^2\big)
\eeq
reproduces resonance saturation for $C_2^{\rm r}(\mu)$
\beq
-\frac{3e}{64\pi^2F_{3\pi}}\frac{1}{\pi}\int^\infty_{4\mpi^2}\diff s' \frac{\Im f_1(s')}{s'^2}
=-\frac{3e}{128\pi^2M_\rho^2}.
\eeq

Besides the issues with the dispersive representation for off-shell kinematics, the off-shell ChPT amplitude itself will in general depend on the parameterization chosen for the pion fields. However, for the present purpose, such effects can simply be absorbed into the definition of $\G$ and $\G'$, whose cancellation in the final expressions therefore ensures that Eq.~\eqref{renormalization} remains reparameterization invariant. 

All these subtleties regarding the off-shell matching between dispersive and chiral amplitudes lead us to the conclusion that the quark-mass renormalization cannot simply be derived from the sum rules in Ref.~\cite{Hannah}.

\end{document}